\documentclass{sig-alternate-nocopy}

\newcommand{\sst}{\phi^{(sst)}}
\newcommand{\sstk}{\phi^{(sst)}_k}

\newcommand{\bibpath}{.}
\newcommand{\figpath}{.}

\begin{document}

\title{{\ttlit{In Silico}} Synchronization of Cellular Populations Through Expression Data Deconvolution}

\numberofauthors{3} 
\author{
\alignauthor
Marisa C. Eisenberg\\
      \affaddr{Mathematical Biosciences Institute}\\
      \affaddr{The Ohio State University}\\
      \email{meisenberg@mbi.osu.edu}
\alignauthor
Joshua N. Ash\\
      \affaddr{Department of Electrical and Computer Engineering}\\
      \affaddr{The Ohio State University}\\
      \email{ashj@ece.osu.edu}
\alignauthor 
Dan Siegal-Gaskins\\
      \affaddr{Mathematical Biosciences Institute}\\
      \affaddr{The Ohio State University}\\
      \email{dsg@mbi.osu.edu}
}

\maketitle
\begin{abstract}
Cellular populations are typically heterogenous collections of cells at different points in their respective cell cycles, each with a cell cycle time that varies from individual to individual.  As a result, true single-cell behavior, particularly that which is cell-cycle--dependent, is often obscured in population-level (averaged) measurements.  We have developed a simple deconvolution method that can be used to remove the effects of asynchronous variability from population-level time-series data.  In this paper, we summarize some recent progress in the development and application of our approach,  and provide technical updates that result in increased biological fidelity.  We also explore several preliminary validation results and discuss several ongoing applications that highlight the method's usefulness for estimating parameters in differential equation models of single-cell gene regulation.
\end{abstract}




\section{Introduction}
In the study of any biological phenomenon, it is important to note that differences between data measured at the level of cellular populations and those collected using single cells can be significant.  This is particularly true for processes that are cell-cycle--dependent, due to the fact that at any given time individual cells within the population can be found at variable points in their respective cell cycles and contribute differently to the population average as a result.   We refer to this kind of variability as \emph{asynchronous variability}, a `feature' of the population that exists even in the absence of cell-cell communication and similar `social' effects, and is independent of any stochasticity in the observable of interest. 

We have developed a simple deconvolution method that can be used to remove the effects of asynchronous variability from population-level time-series data \cite{SiegalGaskins:2009p645}.    An essential component of this technique is an accurate model of the population asynchrony, that is, the temporal position of cells within their cell cycles (their \emph{phase}) and their distribution in the population as a function of experiment time.    The population asynchrony is organism-specific (and possibly condition-dependent as well), and although it may be difficult to establish experimentally, it is in principle characterizable for any system of interest.

\begin{figure}
\centering
\includegraphics[width=0.45\textwidth]{\figpath/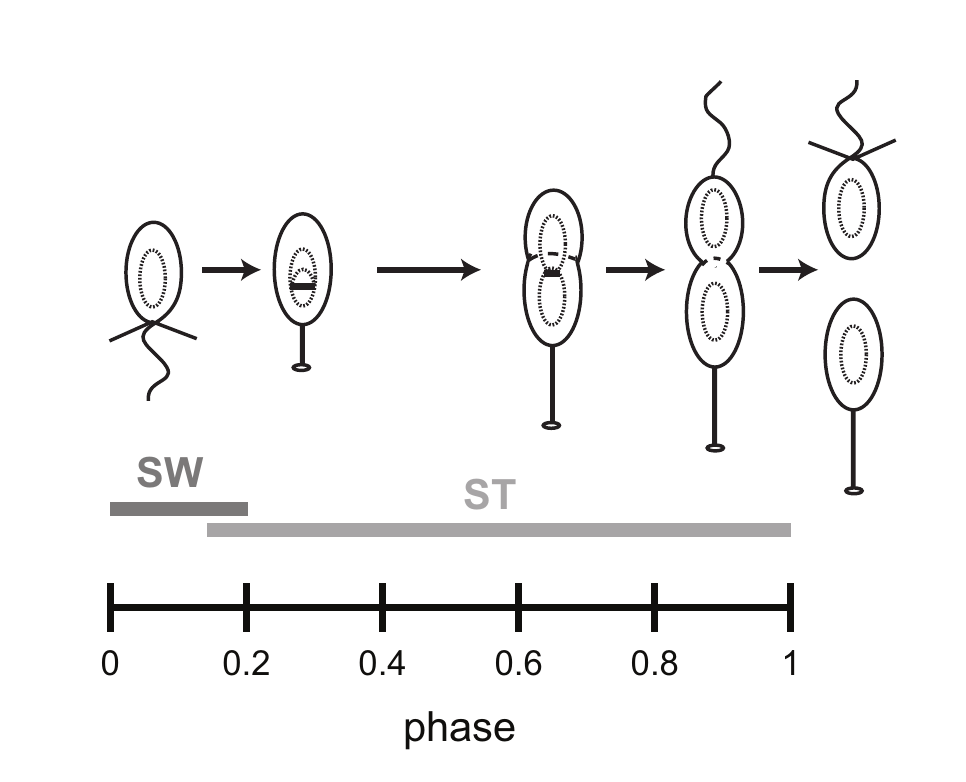}
\caption{\textit{Caulobacter} cell cycle shown with phase axis.  \textit{Caulobacter} begins its cycle as a motile `swarmer' (SW) and differentiates to a non-motile `stalked' (ST) state. Division produces two morphologically distinct cells. }
\label{fig:cellcycle}
\end{figure}

In our previous work \cite{SiegalGaskins:2009p645} we presented an experimentally-validated model for the asynchrony in a population of \emph{Caulo\-bacter crescentus} cells.  \emph{Caulobacter} is a dimorphic bacterium that has been established in both experimental and computational biology communities as an important model organism for the study of cell cycle regulation, cellular differentiation, and the mechanisms of bacterial chromosome segregation \cite{Keiler:2003p770, Li:2009p714, Shebelut:2009p1112}.   The \emph{Caulobacter} cell cycle, which contains both a motile `swarmer' (SW)  stage and a non-motile `stalked' (ST) stage, is shown in Figure \ref{fig:cellcycle}. By applying our deconvolution technique to expression data for a set of \emph{Caulobacter} genes involved in regulating the cell cycle, we showed how critical details obscured in the raw population-based measurements may be recovered.  

We now extend our deconvolution method and demonstrate explicitly, using a `toy' oscillator described by a set of ordinary differential equations, how deconvolution can be used to reconcile population-level expression data with single cell--level mathematical models of gene regulation.    The essentials of the method are summarized below, and we refer the reader to \cite{SiegalGaskins:2009p645} for further details.

\section{Summary of the deconvolution method}
\subsection{Cell cycle phase distribution} \label{sec:cellcycle}  We refer to the position of a cell within its own cell cycle as the its {\it phase} $\phi$, where $0 \leq \phi \leq 1$.   The phase at which an individual cell (indexed $k$) transitions from SW to ST is $\phi^{(sst)}_k$, a random variable normally-distributed in the population with mean $\mu_{sst}$=0.15 (updated from the previous value of 0.25 with new experimental evidence \cite{SiegalGaskins:2009p645, Keiler:2003p770}) and CV=0.13 \cite{SiegalGaskins:2009p645}.  At the outset of a typical batch-culture \emph{Caulobacter} experiment, each cell in the population can be found with $\phi_k(0) \leq \phi^{(sst)}_k$ \cite{MEvinger10011977}.  As experiment time $t$ passes, cells progress through their cycles at a rate that is the inverse of their total cycle times; that is, $\phi_k(t) = \phi_k(0) + t / T_k$ for $0 \leq t \leq T_k (1 - \phi_k(0))$, where $T_k$ is the total cycle time of cell $k$.   When $t = T_k (1 - \phi_k(0))$ and the cell reaches the end of its cycle, two daughter cells---one an SW cell and one an ST cell---emerge.   This phase evolution model can be simulated for a large number of cells and accurately predicts the time-dependent cell cycle phase distribution of an initially all-SW \emph{Caulobacter} culture (see Section \ref{sec:celldist} below).

\subsection{Average expression in single cells and at the population-level} The signal intensity of a particular species measured in a typical RNA expression assay is proportional to the population-level concentration $G(t)=R(t)/V(t)$, where $R(t)$ is the number of transcripts in the population and $V(t)$ is the total cellular volume.  For a large number of cells $N(t)$, 
\begin{equation} 
V(t)  \approx N(t) \int \tilde{Q}(\phi, t) \, d\phi
\end{equation}
and
\begin{equation}
R(t) \approx N(t) \int f(\phi) \tilde{Q}(\phi, t) \, d\phi  \,
\end{equation} 
where $\tilde{Q}(\phi, t)$ is the expected value of a single cell's volume $v_k(\phi)$ over $\theta_k = \{\sstk, T_k\}$, and $f(\phi)$ is  the average expression of all cells at the exact same phase for a given mRNA species, i.e., the fully synchronized average expression. (See \cite{SiegalGaskins:2009p645} for the complete derivation.)  We may thus write the total concentration of RNA transcripts for a particular gene at time $t$ as an integral transform 
\begin{align} 
G(t) &= \frac{R(t)}{V(t)} \nonumber \\ 
&= \frac{ \int f(\phi) \tilde{Q}(\phi, t) \, d\phi}{\int \tilde{Q}(\phi, t) \, d\phi} \nonumber \\ 
&= \int Q(\phi, t) f(\phi) \, d\phi \ , 
\label{eq:intQf}
\end{align} 
where $Q(\phi, t) = \tilde{Q}(\phi, t) / \int \tilde{Q}(\tilde{\phi}, t) \, d\tilde{\phi}$ is the kernel of the transform, and
has the interpretation of a fractional volume density.  That is, $Q(\phi,t)$ represents the fraction of the
total population volume at time $t$ that exists in (a small interval around) phase $\phi$.  
Because of the complexity introduced by cells going through their cycles at different rates (and, in the case of \emph{Caulobacter}, producing new daughter cells at different phases), the deconvolution method relies on simulation methods to evaluate $\tilde{Q}(\phi,t)$ and $Q(\phi,t)$.

\subsection{Estimating single cell expressions from \\population data}\label{sec:singcelexp}Extracting a single cell expression profile from population data involves solving the integral equation \eqref{eq:intQf} for $f(\phi)$ from a limited set of $N_m$ population measurements $\{G(t_1) \ldots$ $G(t_{N_m}) \}$, taken at times $t_1, \ldots, t_{N_m}$.  As $N_m$ is finite and small, this inversion process is ill-posed and requires a degree of regularization; that is, the introduction of additional information in order to reduce the degrees of freedom in the reconstruction process.  To this end, we model $f(\phi)$ as a natural cubic spline
\begin{align}
 f_\alpha(\phi) = \sum_{i=1}^{N_c} \alpha_i \psi_i(\phi) \ ,
 \label{eq.spline_decomp}
\end{align}
where $\{\psi_i(\phi)\}$ represent $N_c$ basis functions, each piecewise cubic polynomials, and $\{\alpha_i\}$ are unknown coefficients that determine the particular shape of the single cell expression.

An optimization problem is solved in order to find coefficients that fit the data well while maintaining smoothness consistent with natural expressions and not over-fitting the measurements.  The cost criterion to be optimized is
\begin{align}
 C(\lambda) = \sum_{m=1}^{N_m} \frac{(G(t_m) - \widehat{G}(t_m))^2}{\sigma_m^2} +
 \lambda \int \{f''(\phi)^2\} d\phi \ ,
 \label{eq.cost}
\end{align}
where $\widehat{G}(t_m) = \int Q(\phi, t_m) f_\alpha(\phi) d \phi$ is the model-predicted measurement and $\sigma_m$ is the variance of the $m$th measurement.  The second term in \eqref{eq.cost} is a second derivative regularization function that encourages smoothness in the estimate.  The parameter $\lambda$ controls the tradeoff between data fidelity and smoothness and may be selected via cross validation \cite{Craven:1979, SiegalGaskins:2009p645}.

In addition to smoothness, a number of other physically based constraints are included to increase the fidelity of the model and improve estimation performance.  These include
\begin{enumerate}
 \item \textit{Positivity.} As the expression cannot be negative, we impose the constraint that $f_\alpha(\phi)$ cannot be negative. \\
 \item \textit{Continuity.} 
The expression is also constrained by the fact that we must have conservation of RNA species across cell division.  In other words, the concentration of any RNA species at phase $\phi = 1$ must be equal to the volume-weighted sum of concentrations at phases $\phi = 0$ and $\phi = \sst_k$ for every cell $k$.  This constraint can be modeled as $\int w(\phi) f_\alpha(\phi) d \phi = 0$, where the form of $w(\phi)$ depends on the particular cell division model (see \cite{SiegalGaskins:2009p645}).
\end{enumerate}

The single cell estimate $f_\alpha(\phi)$ is determined by the set of $\alpha$-coefficients that minimize \eqref{eq.cost} while satisfying all of the constraints above.

\section{Method updates}
In this section we present two refinements to the deconvolution method presented in \cite{SiegalGaskins:2009p645}. The first proposes a more realistic cell volume model where changes during growth and division occur smoothly. Similarly, the second constrains the resultant expression estimate to enforce continuity in the rate of transcription generation across cell division.

\subsection{Smoothness of cell volume function}\label{sec:cellvol}
The integration kernel $Q(\phi, t)$ depends on $\tilde{Q}(\phi, t)$, which in turn relies on a cell volume model $v_k(\phi)$ describing the volume of the $k$th cell as a function of its phase $\phi$ and parameters $\theta_k$.  It has previously been shown that, on cell division, the \emph{Caulobacter} cell volume is partitioned 40\% SW  to 60\% ST \cite{Thanbichler:2006p271}.  This implies that
\begin{align}
 v_k(1) &= V_0 \label{eq.v1}\\
 v_k(0) &= 0.4 \, V_0 \\
 v_k(\sstk) &= 0.6 \, V_0 \ ,
\end{align}
where $V_0$ is the cell volume at $\phi=1$, just prior to division.  Here we assume that the variance of the final cell size distribution is small so that $V_0$ is effectively constant across all cells $k$.  Further, the rate of volume change of the two cell halves should be continuous across division
\begin{align}
 v_k'(0) &= v_k'(1)\\
 v_k'(\sstk) &= v_k'(1) \ , \label{eq.v5}
\end{align}
that is, both the swarmer cell and stalked cell should start with the same rate of volume change as the cell just prior to division.  We adopt a simple piecewise polynomial model that satisfies \eqref{eq.v1}--\eqref{eq.v5} and is consistent with known biology:

$\displaystyle v_k(\phi) =  $
\begin{equation}
V_0 \times\left\{
 \begin{array}{ll}
 0.4 + \frac{0.4}{1 - \sstk} \phi & \\
 + \frac{0.6 - 1.8 \sstk}{(1 - \sstk) (\sstk)^2} \phi^2 , & 0 \leq \phi < \sstk\\
 + \frac{1.2 \sstk - 0.4}{(1 - \sstk) (\sstk)^3} \phi^3 &  \\
 \\
 1 - \frac{0.4}{1 - \sstk} + \frac{0.4}{1 - \sstk} \phi , & \sstk \leq \phi < 1
 \end{array}
 \right.  \label{eq:newvolume}
\end{equation}
The second piece is linear, and the first piece is approximately linear with small quadratic and cubic terms.  This model extends the purely linear model in \cite{SiegalGaskins:2009p645} which did not consider the volume rate-of-change constraints.

\subsection{Smoothness of gene expression}
In addition to the continuity constraint described in Section \ref{sec:singcelexp}, we also expect the rate of change of number of RNA transcripts to also be continuous across cell division.  Intuitively, cells should not, for example, have net transcript creation at a given phase and instantaneously switch to net transcript degradation.  Mathematically, if $R_k(\phi) = v_k(\phi) f_k(\phi)$ represents the number of transcripts from cell $k$, the constraint may be written in terms of the derivative with respect to cell phase, $R'_k(1) = R'_k(0) + R'_k(\sst_k)$.  Using the chain rule for differentiation along with the properties and form of the cell volume function in the previous section, the constraint may be written
\begin{align}
	\begin{split}
			\beta(\sstk) (f_k(1) - f_k(0) - f_k(\sstk)) = \\ 
	     0.4 f'_k(0) + 0.6 f'_k(\sstk) - f'_k(1) ,
	\end{split}
	\label{eq:vfderiv}
\end{align}
where $\beta(\sstk) = v'_k(1)/V_0 = 0.4/(1 - \sst_k)$.  Taking the synchronous average, denoted $\langle \cdot \rangle$, of \eqref{eq:vfderiv} over a large number $N$ of cells yields
\begin{align}
	\begin{split}
			\langle \beta(\sstk) \rangle f(1)-& \\
			\langle \beta(\sstk) &\rangle f(0) - \langle \beta(\sstk) f_k(\sstk \rangle = \\
			&0.4 f'(0) + 0.6 \langle f_k'(\sstk) \rangle - f'(1)
	\end{split}
	\label{eq.sync_constraint}
\end{align}
where $f(\phi) = \langle f_k(\phi) \rangle$ is the synchronous average expression.  Noting that the variability between the $f_k$ is independent of $\sstk$, the averages in \eqref{eq.sync_constraint} simplify to
\begin{align}
			\langle \beta(\sstk) \rangle & = \frac{1}{N} \sum_k \beta(\sstk) \nonumber \\
                                  &\approx \int \beta(\phi) p(\phi) \, d \phi \nonumber \\
                                  &= \beta_0
			\label{eq.avg1}
\end{align}
\begin{align}
			\langle f_k'(\sstk)  \rangle & = \frac{1}{N} \sum_k f_k'(\sstk) \nonumber \\
                                  & \approx \frac{1}{N} \sum_k f'(\sstk) \nonumber \\
                                  & \approx \int f'(\phi) p(\phi) \, d \phi,
			\label{eq.avg2}
\end{align}
and
\begin{align}
			\langle \beta(\sstk) f_k'(\sstk)  \rangle & = \frac{1}{N} \sum_k \beta(\sstk) f_k(\sstk) \nonumber \\
                                               & \approx \frac{1}{N} \sum_k \beta(\sstk) f(\sstk) \nonumber \\
                                               & \approx \int \beta(\phi) f(\phi) p(\phi) d \phi,
			\label{eq.avg3}
\end{align}
where $p(\phi) = \mathcal{N}(\phi; \mu_{sst}, \sigma^2_{sst})$ is the Gaussian probability density function of $\sstk$. Substituting \eqref{eq.avg1}--\eqref{eq.avg3} into \eqref{eq.sync_constraint} provides the final constraint
\begin{align}
	\int w_1(\phi) f(\phi) d \phi = \int w_2(\phi) f'(\phi) d \phi,
	\label{eq.derivative_constraint}
\end{align}
where
\begin{align}
	w_1(\phi) &= \beta_0 \delta(1 - \phi) - \beta_0 \delta(\phi) - \beta(\phi) p(\phi)   \\
	w_2(\phi) &= 0.4 \delta(\phi) + 0.6 p(\phi) - \delta(1-\phi),
\end{align}
and $\delta(\phi)$ is the dirac delta function.  
\\
\\
\\

\begin{figure}
\centering
\includegraphics[width=0.42\textwidth]{\figpath/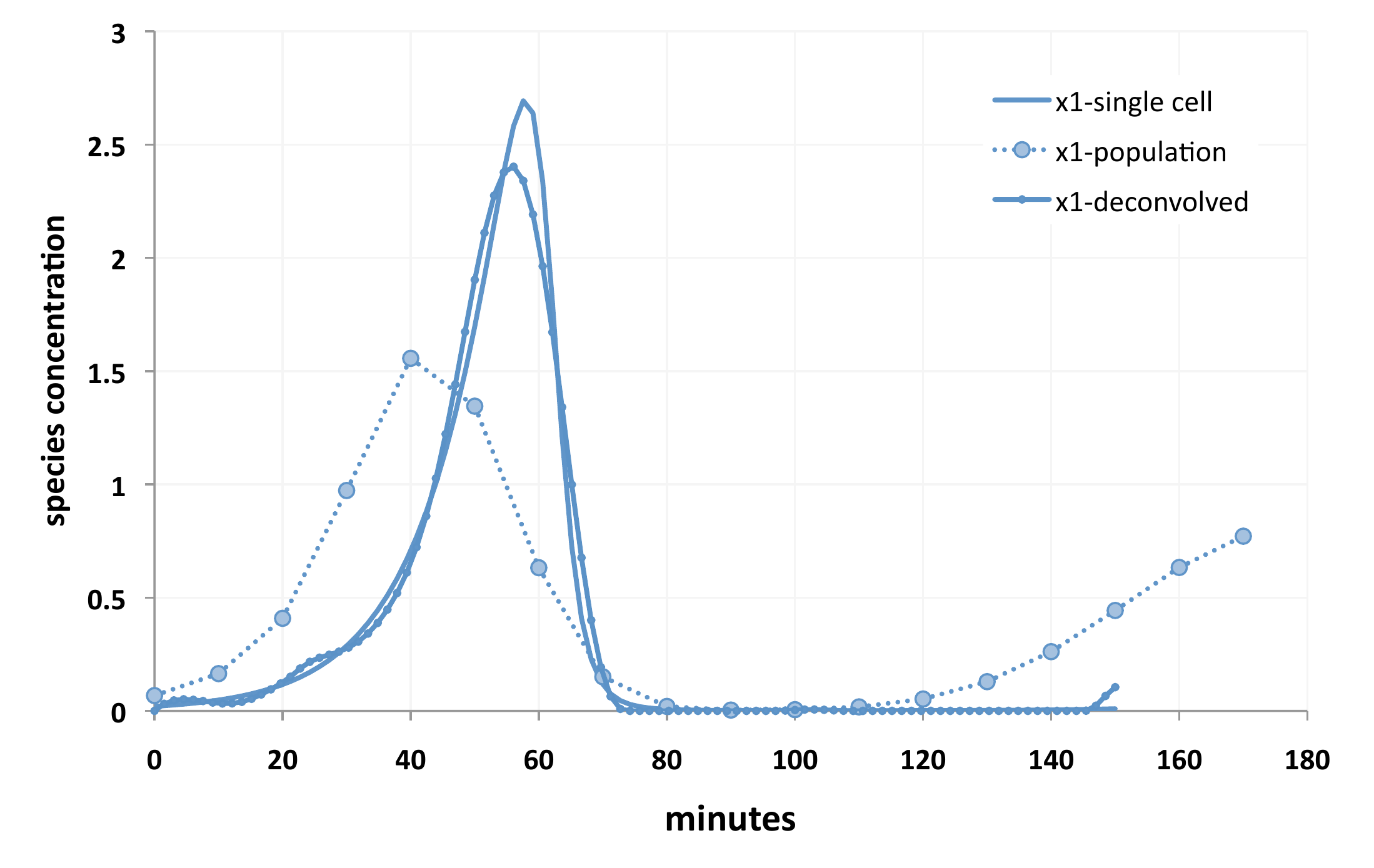}\\
\includegraphics[width=0.42\textwidth]{\figpath/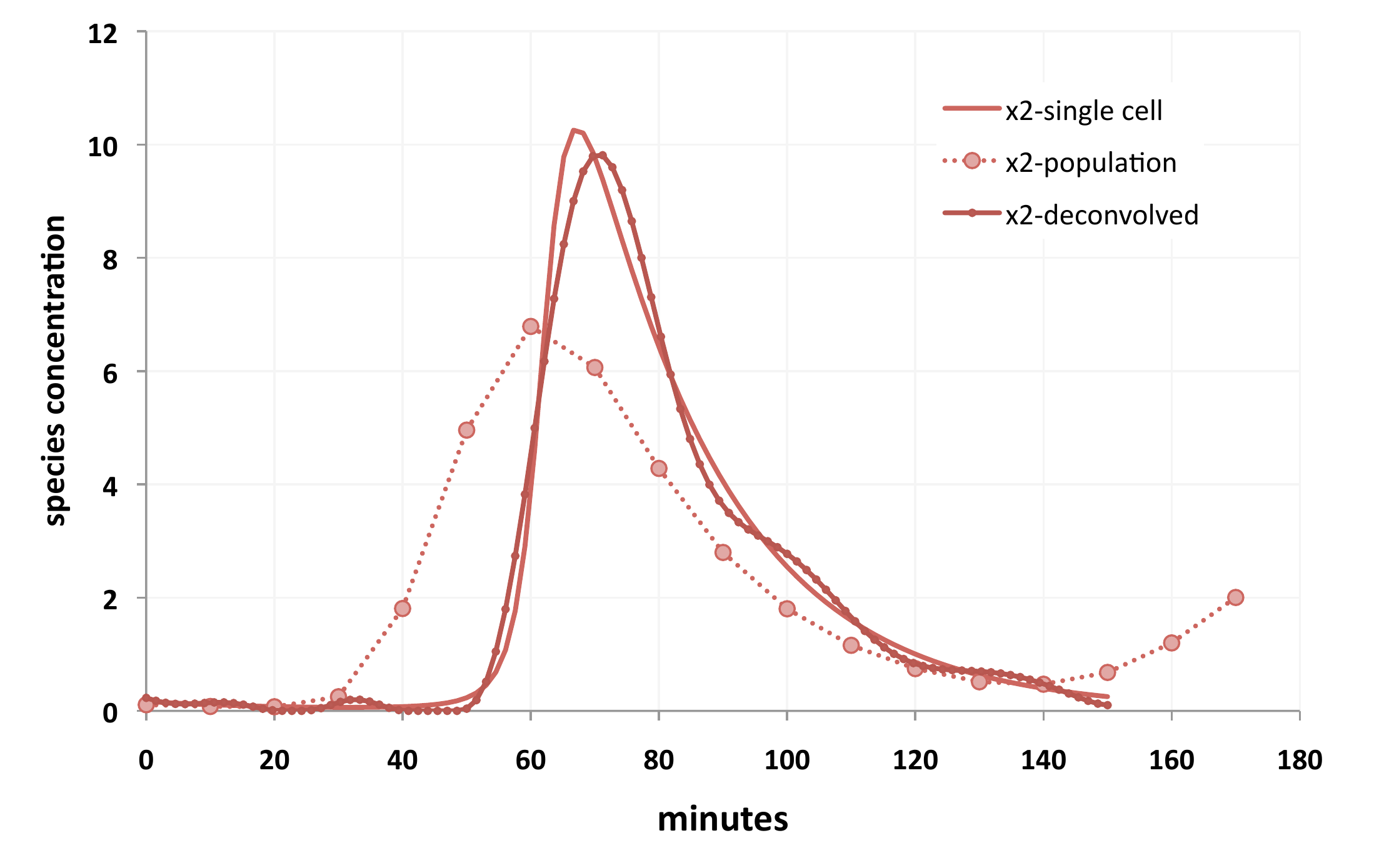}
\caption{`True' synchronized single cell simulations of a simple biological oscillator model, compared to the resulting population and deconvolved expressions.}
\label{fig:LVnoiseless}
\end{figure}

\section{Validation \& Applications}
\subsection{Biological oscillator model deconvolution}
In this section we explore the performance of the deconvolution process by testing it on simulated population data, where the `true' synchronized cell behavior is known.  Here a particular model of a cell-cycle regulated expression in single cells is passed through the forward model using the kernel function $Q(\phi,t)$ in order to generate simulated population-level data.  

As an example differential equation system, we consider the classical Lotka-Volterra model \cite{lotkavolterra} as a biological oscillator.  Although originally developed to represent predator prey dynamics, the Lotka-Volterra equations have also been applied in a regulatory network context (see, e.g., \cite{Boys:2008p1114}), making them a useful test case.  The model equations are given by:
\begin{align}
\dot{x_1} = x_1 (a - b x_2)\\
\dot{x_2} = x_2 (c x_1 - d)
\end{align}
Where we take $x_1$ and $x_2$ to represent two chemical species which bind and convert $x_1$ to $x_2$.  

We chose parameter values which yield a 150 minute period oscillation (similar to the average cell cycle time for \emph{Caulobacter}), and convolved the resulting simulations with $Q(\phi,t)$ to yield a noiseless asynchronous population.  We then added a several of levels and types of noise to the population data, and deconvolved the simulated population data to test how well the method recovers the known `single-cell'--like behavior.  

\begin{figure}
\centering
\includegraphics[width=0.42\textwidth]{\figpath/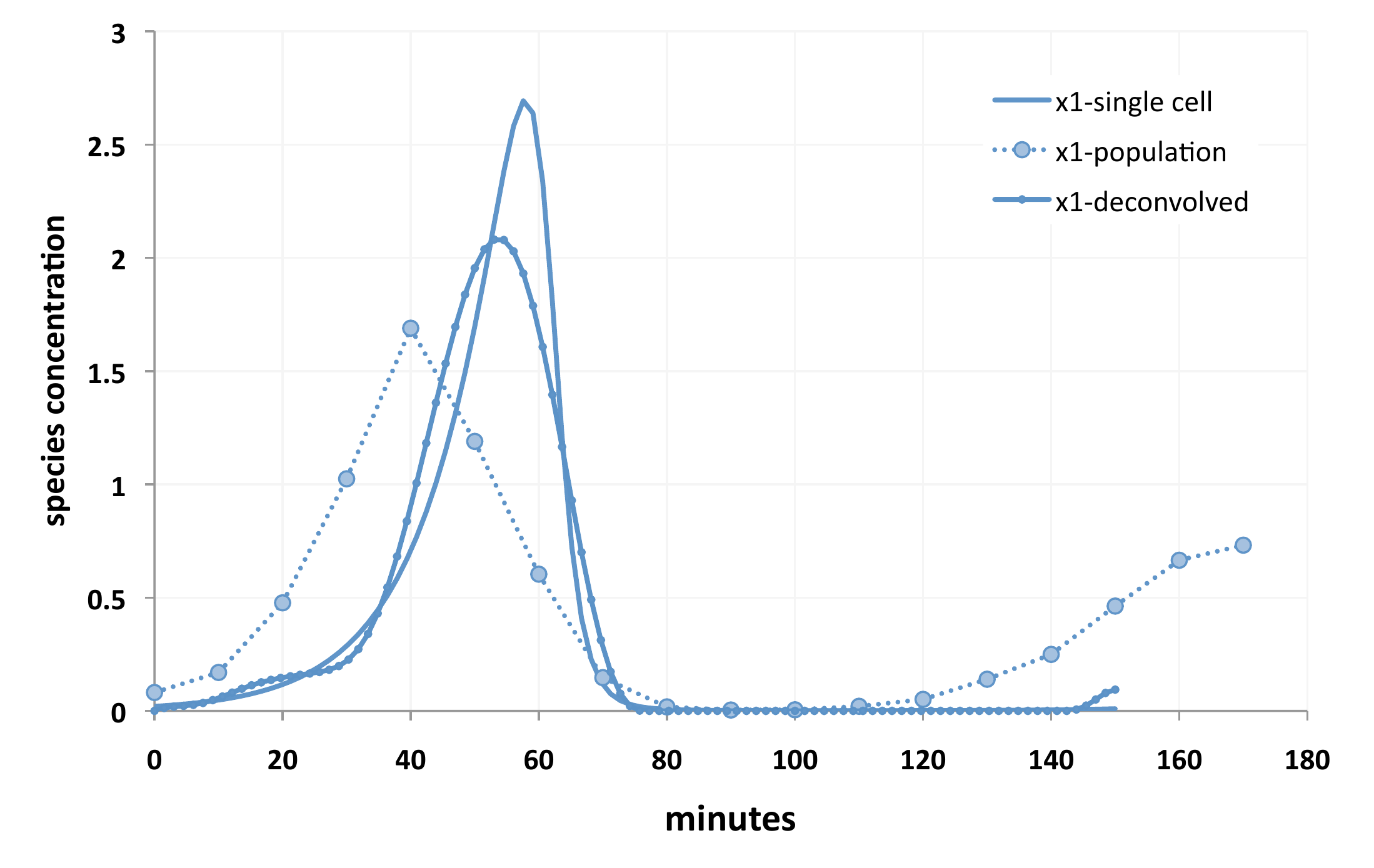}\\
\includegraphics[width=0.42\textwidth]{\figpath/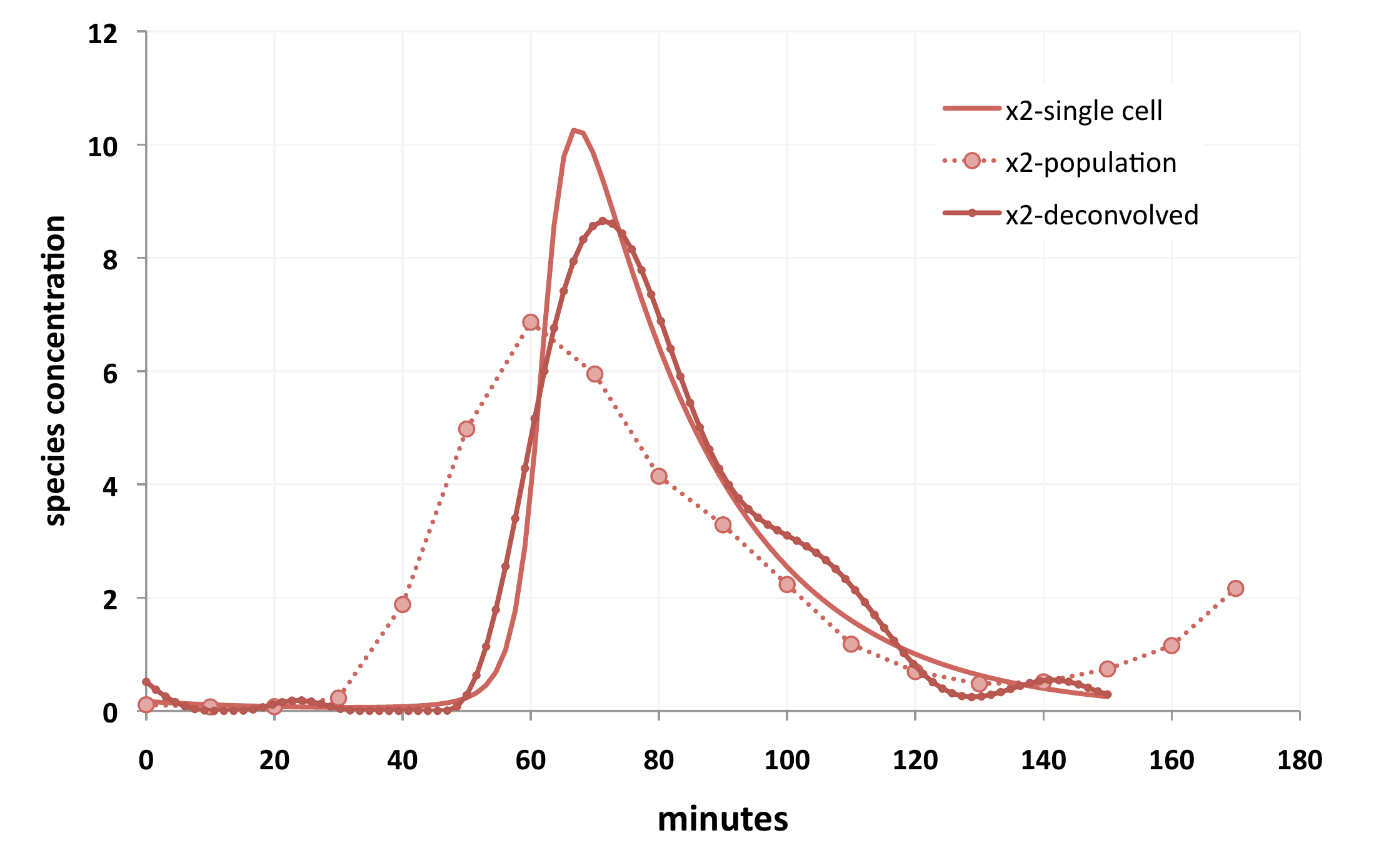}
\caption{`True' synchronized single cell simulations of a simple biological oscillator model, compared to one realization of the resulting population and deconvolved expressions, where Gaussian error with standard deviations equal to 10\% of the data magnitude have been added to the population data.}
\label{fig:LV10percent}
\end{figure}

Figures \ref{fig:LVnoiseless} and \ref{fig:LV10percent} show two examples 
from these investigations for the noiseless case and a case where Gaussian distributed errors with mean zero and standard deviation equal to 10\% of the data magnitude have been added to the population data.  The deconvolution generally performs well at recovering the major features of the synchronous cell behavior, and these results suggest that this method may be a useful tool for parameter estimation for models developed to represent gene regulation in individual cells.

\subsection{Distribution of cell types in \secit{Caulobacter}} \label{sec:celldist} 
\begin{figure}
\centering
\includegraphics[width=0.4\textwidth]{\figpath/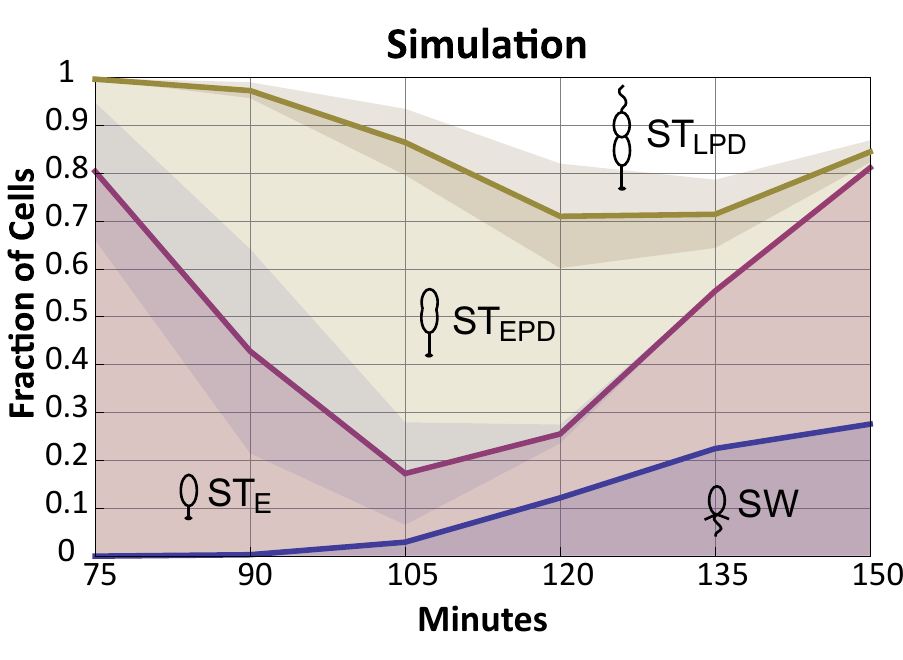}
\includegraphics[width=0.4\textwidth]{\figpath/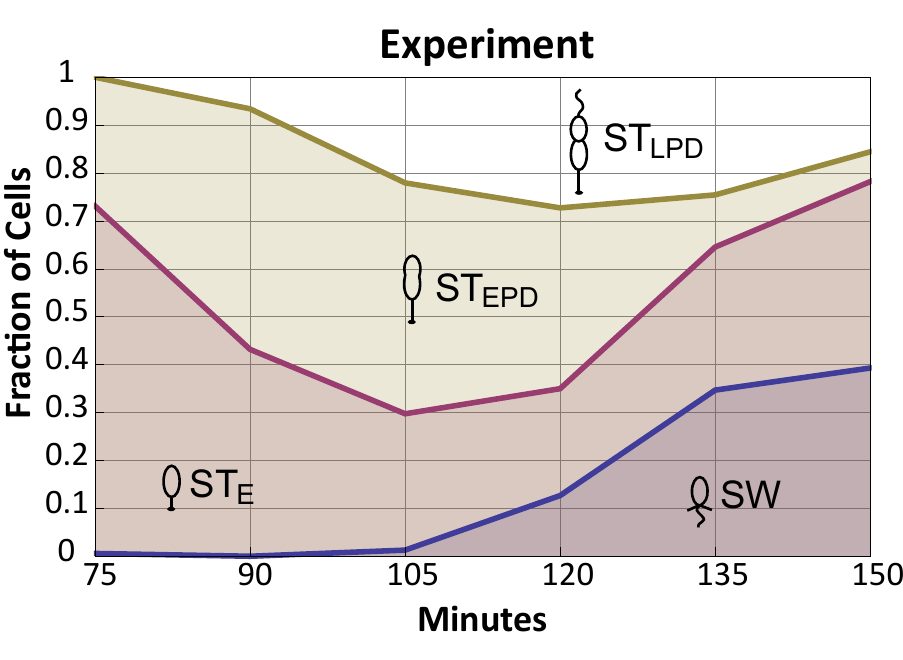}
\caption{The simulated distribution (top panel) of a batch-culture population of \emph{Caulobacter} matches the experimentally-observed distribution (bottom panel). Experimental data is reproduced from Judd et al. \cite{Judd08072003}. Shaded regions in the simulated distribution indicate a range of transition phases, with the solid line indicating the midpoint.}
\label{fig:celldist}
\end{figure}

Using the cell-type distribution model presented here, we can simulate the distribution of cell types over time in a typical batch-culture \emph{Caulobacter} experiment, and compare this to experimental data on cell type distributions.  Simulated cells were grouped based on their cell cycle phase into swarmer (SW) and stalked (ST), with the ST cells split into early stalked (ST$_E$), early predivisional (ST$_{EPD}$), and late predivisional (ST$_{LPD}$).  We assume a mean SW-ST$_E$ transition phase of 0.15 as discussed in Section \ref{sec:cellcycle}.  For the ST$_E$-ST$_{EPD}$ and ST$_{EPD}$-ST$_{LPD}$ transition phases, we note that distinguishing between ST$_E$ and ST$_{EPD}$, and ST$_{EPD}$ and ST$_{LPD}$ morphologies can be difficult experimentally, so we use a range of  0.6-0.7 for  ST$_E$-ST$_{EPD}$ and a range of 0.85-0.9 for ST$_{EPD}$-ST$_{LPD}$, based on \cite{Judd08072003, SiegalGaskins:2009p645}.  The resulting time dependent cell type distributions, compared to experimental data from Judd et al. \cite{Judd08072003}, are shown in Figure \ref{fig:celldist}.  Our cell-type distribution model predicts highly similar distributions of each cell type, providing additional experimental validation support for our cell type model.

\subsection{{\secit ftsZ} expression in \secit Caulobacter}  
\begin{figure}
\centering
\includegraphics[width=0.4\textwidth]{\figpath/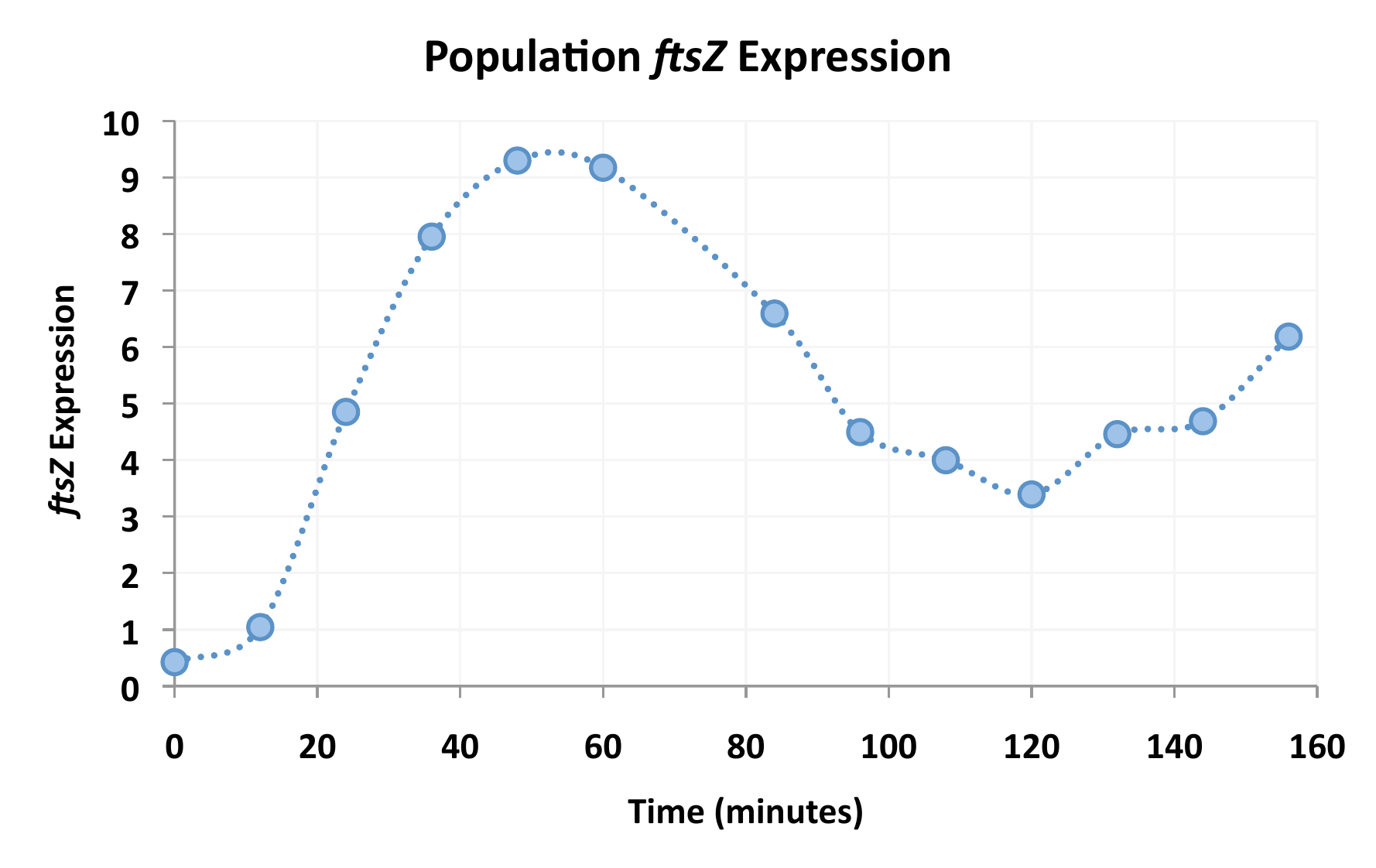}\\
\includegraphics[width=0.4\textwidth]{\figpath/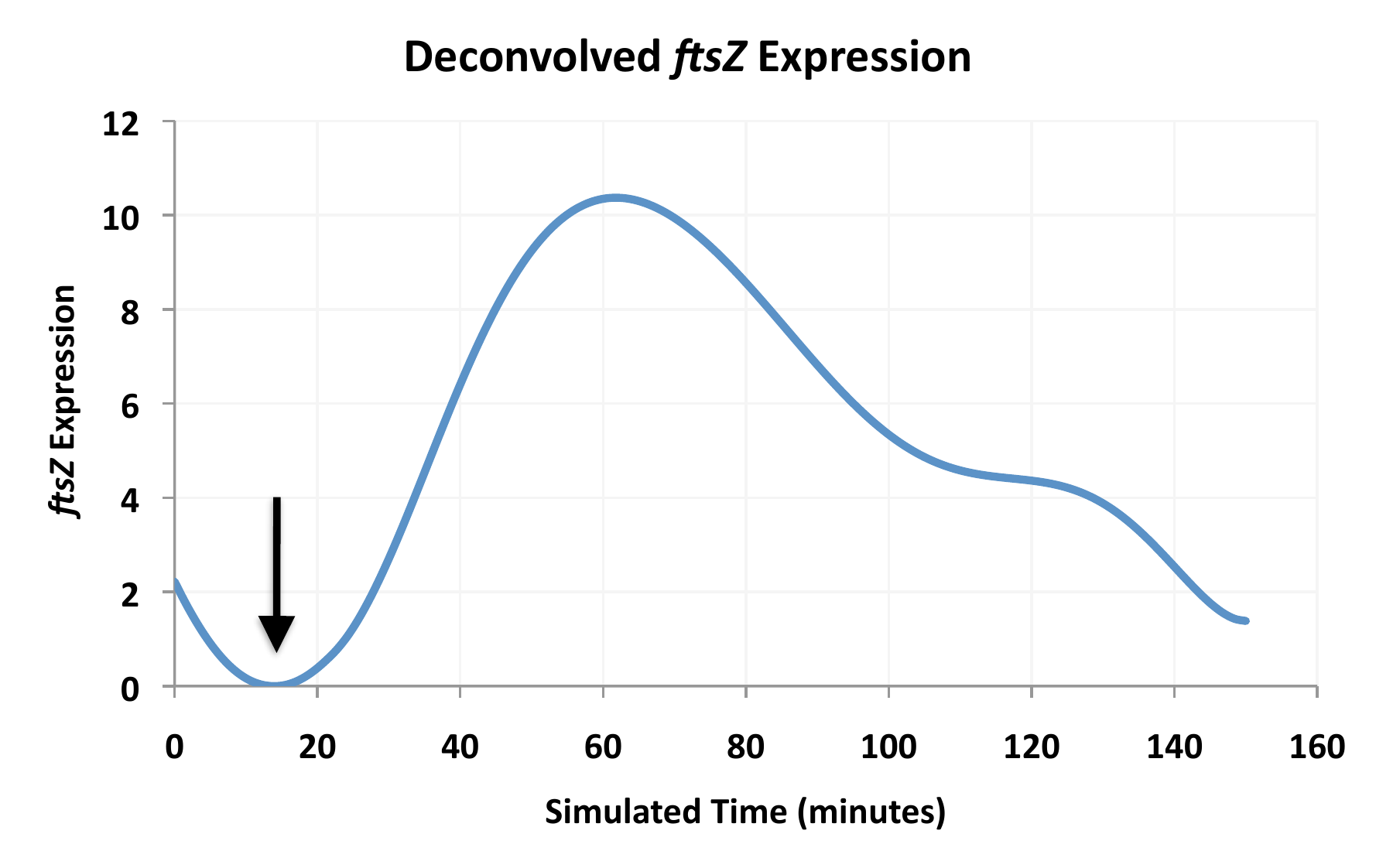}
\caption{Population vs. deconvolved \emph{ftsZ} expression in \emph{Caulobacter}.  The deconvolved data (bottom panel) resolves the \emph{ftsZ} transcription delay (indicated by the arrow) not observed in the population data (top panel).  Simulated time is a scaling of the cell cycle phase to the average cell cycle time of 150 minutes.}
\label{fig:ftsZ}
\end{figure}

To highlight the usefulness of our method with a concrete example, we show here the population-level and deconvolved data for the \emph{Caulobacter} \emph{ftsZ} gene. (Population data was taken from \cite{Mcgrath:2007}.)   FtsZ is a tubulin homolog essential for bacterial cell division that is produced only after DNA replication begins at the SW-to-ST transition \cite{Kelly:1998p282}.   This delay in \emph{ftsZ} transcription cannot be seen in the microarray data, but is visible in the deconvolved expression profile (Figure \ref{fig:ftsZ}), providing validation for the deconvolution method. The deconvolution makes an additional new prediction of a large drop in the level of \emph{ftsZ} with no subsequent increase after transcript concentration reaches its maximal value at $\phi \approx 0.4$.   This deviates significantly from the raw microarray data, which shows levels of \emph{ftsZ} increasing towards the end of the experiment.

\section{Conclusions \& ongoing work}
To conclude, we have developed a method for using deconvolution to synchronize populations of cells \emph{in silico}, which yields expression information that is more representative of the `true' single-cell expression.  Here, we presented three updates to this method: an updated SW-ST transition phase, an additional smoothness condition, and an updated cell volume function.  We also present new validation and applications using experimental data as well as a differential equation model of a simple biological oscillator.  We are currently extending this work to explore the applications of this method in estimating parameters for differential equation models of gene regulatory networks, which are typically built to model single cell behavior but fitted to population data.  Our ongoing results suggest that the deconvolution technique and asynchronous cell population model yield more accurate single cell parameters than fitting to population data alone.

\section{Acknowledgments} 
This work was partially supported by the National Science Foundation (Agreement 0635561) and an NIH T32 training grant from the Division of Human Cancer Genetics to DSG.

\balancecolumns

\bibliographystyle{abbrv}
\bibliography{\bibpath/bibliography.bib}  

\end{document}